# Stress Driven Morphological Instabilities in Rocks, Glass, and Ceramics

M.A. Grinfeld,
US Army Research Laboratory
Aberdeen Proving Ground, MD, 21005-5069

Abstract. The purpose of the study is to further investigate the classical Gibbs analysis of the heterogeneous system "stressed crystal – melt." It is demonstrated that each equilibrium configuration is stable with respect to a special class of variations introduced by Gibbs. This basic result is compared with the opposite result on the universal morphological instability of phase interface separating a stressed crystal with its melt. Some plausible manifestations of the instabilities implied by the Gibbs model are qualitatively discussed.

## *Introduction*

Almost 150 years ago, Gibbs suggested his solution to the problem of equilibrium of nonhydrostatically stressed solid existing in contact with its solution [1]. Gibbs analyzed this problem by means of his famous variational paradigm. This is the key problem of geophysics, volcanology, metallurgy, and materials science, and it was analyzed by many remarkable physicists, including Bridgman [2], Lifshitz [3], Nozieres [4], and others. A more detailed historical review can be found in the monograph [5]. Despite many efforts, researchers in different disciplines are still very far from reaching a consensus, and an endless flux of severe criticisms and controversies can be found in the literature. Current scientific literature is full of mutual criticisms, and we are still very far from a clear understanding of this inexhaustible problem in which so many things remain to be carefully investigated. Presumably, we are still very far from the level of understanding of this problem reached by Gibbs himself. Numerous worldwide practitioners have to use various existing interpretations of this key problem, possibly, thousands of times a day, seven days a week. This makes it necessary to brush up on our understanding, to return back to the classical analysis of Gibbs and to realize how far we deviated from his vision.

Theoretical interest of the Gibbs analysis has grown considerably after the discovery of rather universal stressed driven rearrangement instabilities (SDRI) of phase interfaces, especially by the SDRI "stressed crystal – melt" (see publications [6–10]). The SDRI, discussed in [5–10], are the direct logical implications of the Gibbs paradigm. Unfortunately, these implications of the Gibbs paradigm obviously contradict numerous everyday observations and precise scientific experiments. However, several plausible physical manifestations of the SDRI have been mentioned in [5,7,9]. Still, none of these SDRI have been reliably confirmed in the experiment (however, see some interesting attempts in [11,12]). Nevertheless, the SDRI caused a lot of enthusiasm in nanophysics, epitaxy, fracture theory, etc. Of course, some of these hopes are the results of biased hopes, as it was clarified by Nozieres [13], who himself is one of the most enthusiastic advocates of the SDRI "stressed crystal – melt." All things considered, the Gibbs theory requires further developments and clarifications.

In this paper, we first restate and comment on the original analysis of the Gibbs gedanken experiment (as this author understands it). Our presentation is somewhat different as compared with [1], although it is still as general as the original Gibbs analysis. First, assume upfront that the system's temperature is spatially uniform throughout the system. Also, we use somewhat different notation which is more convenient for the discussion of stability.

The main new result of this paper is presented in the second section. It is demonstrated that the system "nonhydrostatically stressed crystal – melt" is stable with respect to the set of variations used by Gibbs.

To the best of the author's knowledge, this fact has never before been demonstrated. Moreover, in the current literature there appeared an opinion that the instability in the system "stressed crystal – melt" was demonstrated by Gibbs himself. We then discuss the SDRI "stressed crystal – melt" on the qualitative level and clarify some misinterpretations of the author's earlier publications [5-10].

## *The Gibbs gedanken experiment and his results*

Following the original analysis of Gibbs, the deformable solid is assumed nonlinear and anisotropic. The notation and presentation of the basics in the following, however, differ from the original notation of Gibbs [1]. They are taken from [5], especially, from §5.4 "The problem of equilibrium shape of a deformable elastic crystal." The crystalline elastic substance is referred to the Lagrangean coordinates $x_i$, which generate an affine coordinate system in the reference configuration. This coordinate system spreads over the entire space, although the crystal itself occupies only part of this space, and this part varies in the process of dissolution/precipitation. In the following, we get rid of the too elaborate tensorial notation, choose the Cartesian coordinate system in the reference configuration, and do not distinguish between the lower and upper indexes. However, we still adopted the summation rule over repeated indexes. The indexes *a,b,c* run over 1,2, whereas the indexes *i,j,k,l,m,n* run over the values 1,2,3. In particular, the notation $x^i$ symbolizes the set $\{x_1, x_2, x_3\}$ whereas $x^a$ symbolizes the set $\{x_1, x_2\}$. The symbol $\delta_{ij}$ is the standard 3D Kronecker delta, $\delta_{ab}$ is the standard 2D Kronekker delta, and $\delta_{ai}$ is equal to 1 when $a = i$ and it is equal to 0 at $a \neq i$.

The metrics $x_{ij}$ of the reference configuration is equal to $\delta_{ij}$. Let $a_{ij}$ be the deformation tensor (analogous to the nine-element matrix (354) of [1]). Following Gibbs, this tensor is the same for all material points of the crystal. Then, the *actual* covariant metrics is defined as $X_{mn} \equiv a_{im}a_{in}$. The only physically meaningful measure of deformation is the tensor of finite deformations [13] — $u_{ij} \equiv (X_{ij} - x_{ij})/2$ — nonlinearly depending on the displacements.

Let $\psi_c(u_{ij}, \theta)$ be the free energy density per unit mass of the crystalline substance, which is a function of the finite deformations and the absolute temperature $\theta$. Consider a homogeneous liquid solution of $M_l$ mass units of the dissolver and $M_r$ mass unit of the dissolved crystalline substance. Let $\Psi_l(M_l, M_d, V_s, \theta)$ be the total free energy of the solution occupying the volume $V_s$ and kept at the temperature $\theta$; when dealing with a melt instead of a solution, then the total free energy of the melt can be presented as $\Psi_l = M_m \psi_m(\rho_l, \theta)$.

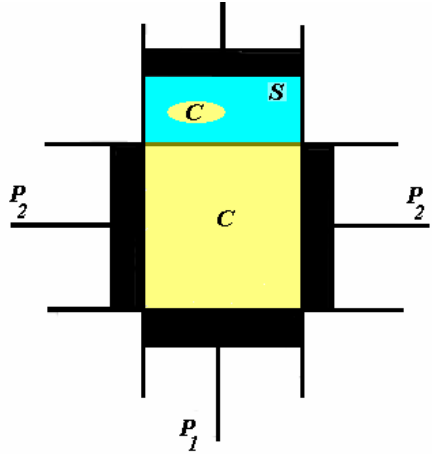
Figure 1. Towards Gibbs gedanken experiment.

The Gibbs experiment is schematically presented in Figure 1. The crystalline parallelepiped (marked with "C") is brought into contact with its liquid solutions (marked with S). Different pressures $P_1$ and $P_2$ are applied to the solutions. As a result, the crystal itself is nonhydrostatically stressed with the maximum shear stress $\tau = |P_1 - P_2|/2$.

For the analysis of equilibrium in heterogeneous systems with mass exchange, Gibbs proposed a variational approach based on the minimum energy principle combined with the second law of thermodynamics. One implication of the Gibbs analysis shows that in each equilibrium configuration the absolute temperature of the system should be spatially uniform. If this fact is assumed upfront then the analysis of Gibbs can be simplified, especially when dealing with the systems kept at constant temperature. In this case, the analysis can be reduced to minimization of the total free energy of the system. This total energy includes three main ingredients:

$$\Psi_{tot} = \Psi_c + \Psi_l + \Psi_\Sigma = \int_{\Omega_c} d\Omega \rho_c \psi_c + \Psi_l + \Psi_\Sigma , \qquad (1.1)$$

where the first and second ingredients are the total bulk energies of the crystalline and melted phases and the last term represents the interfacial energy; $\Omega_c$ and $\rho_c$ are the total spatial volume and mass density of the crystal in the deformed configuration.

The key element of the Gibbs paradigm is the proper choice of admissible variations. When dealing with phase interfaces they should definitely include the possibility of mass exchange between the phases. The set of possible variation also reflects the possibility of independent variation of the displacements of different material points of the crystal. However, in his original analysis Gibbs limits himself with the spatially uniform displacement gradients $a_{ij}$. Usage of this set allows Gibbs to avoid any integration and partial differential equations and deal with the algebraic equation of the total bulk energy

$$\Psi_{bulk} = M_c \psi_c + M_m \psi_m , \qquad (1.2)$$

where $M_c$ and $M_m$ are the total masses of the crystal and the melt or solution.

The Gibbs's original analysis was limited with uniform deformations of the crystals with flat interfaces "crystal – solution." This limitation considerably simplifies his quantitative analysis that deals with nonlinear algebraic equations instead of nonlinear partial differential equations. Other than that, the analysis of Gibbs is practically absolutely universal. In particular, he made no traditional simplifying assumptions about the linearity and/or isotropy of the solid phase. In what follows, our own analysis is limited to the simplest Gibbs setting.

Gibbs demonstrated that, in addition to external boundary conditions and the standard conditions of mechanical and thermal equilibrium, one more equilibrium equation,

$$\psi_c(u_{mn},\theta) + pv_c = \mu_d , \qquad (1.3)$$

should be satisfied in order to guarantee the equilibrium with respect to the process of dissolving/precipitating of the crystalline substance. In the equation (1.3), $\mu_d \equiv \partial \Psi / \partial M_d$ is the chemical potential the dissolved substance, $p = -\partial \Psi / \partial V_s$ is the pressure in the solution and $v_c$ is the actual mass density of the crystalline substance under the action of nonhydrostatic loading. When dealing with melting processes the chemical potential of the melt $\mu_m$ appeared to be equal to $\psi_m + \rho_m \partial \psi_m / \partial \rho_m$.

When $\psi_c + pv_c < \mu_d$ ($\psi_c + pv_c > \mu_d$), the precipitating (dissolving) or crystallizing (melting) occurs until the equilibrium with respect to the mass exchange is restored.

## *The stability of equilibrium configurations with respect to the Gibbs variations*

**Patterns of epitaxial growth.** Keeping in mind further applications to some current applications of the Gibbs analysis, let us dwell on the case of epitaxial crystal growth from melt of vapor. Such systems are showed schematically in Figure 2. A relatively thick monocrystalline substrate (yellow) is placed inside a chamber filled with a vapor (light blue), which then crystallizes on the substrate. The growing monocrystalline film (dark blue) appears to be highly stressed by the mismatch between the lattice parameters and to some extent due to the vapor pressure.

It is known that there are three basic modes of epitaxial growth of solid films: the island (or Volmer-Weber) pattern, the layer-by-layer (or Frank-van der Merwe) pattern, and the mixed (or Stranski-Krastanow) pattern. When the growth follows the island pattern the substrate is covered by the system of separately standing islands that are nonuniformly stressed. When the growth follows the layer-by-layer pattern, the film is uniformly stressed and has a flat interface. These two patterns of growth of solid films are totally analogous to the behavior of liquid substances when they are not or able to wet the substrate.  At last, in the case of the mixed pattern, the growth begins as the layer-by-layer pattern; however, after reaching the critical thickness it proceeds in accordance with to the island pattern. The mixed pattern has no analogies in liquid condensation.

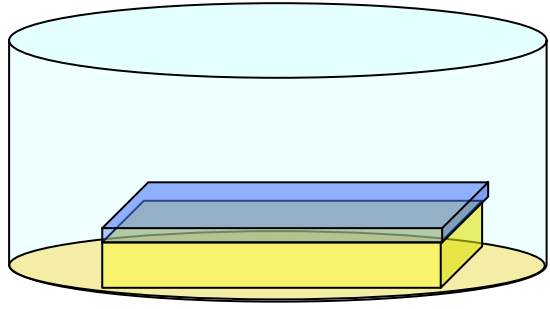

Figure 2. A drawing of epitaxial growth from melt or vapor.

**The Gibbs variations.** The Gibbs original analysis based on the Gibbs uniform variations seems to be an appropriate tool for the analysis of the layer-by-layer pattern. The deformation vector $u^i$ of the uniformly deformed growing film can be presented as follows:

$$u_i = \Gamma_{ab} x_b \delta_{ai} + \gamma_i x_j n_j + v x_j n_j n_i , \qquad (2.1)$$

where $\Gamma_{ab}$ is the in-plane misfit deformation, $\gamma_i$ is the shear vector, and $v$ is the normal extension/contraction. The vector $\gamma_i$ is parallel the coordinate plane $(x_1, x_2)$ and is orthogonal to its unit normal $n_i = \delta_{i3}$. Obviously, any vector $\gamma_i$ can be presented in the form $\gamma_i = \delta_{ai} \gamma_a$, where $\gamma_a$ is an appropriate 2D vector. Thus, the set of the Gibbs variations depends on the three independent parameters: the matrix $\Gamma_{ab}$, the 2D vector $\gamma_a$, and the scalar $v$.

For the displacements (2.1), the deformation gradient matrix $a_{ij}$ and the actual metrics $X_{ij}$ are equal to

$$a_{kl} = \delta_{kl} + \Gamma_{ab} \delta_{ak} \delta_{bl} + \gamma_k n_l + v n_k n_l \qquad (2.2)$$

and

$$X_{mn} - x_{mn} = (\Gamma_{ab} + \Gamma_{ba})\delta_{am}\delta_{bn} + \Gamma_{ab}\Gamma_{ad}\delta_{bm}\delta_{dn} + \gamma_n n_m + \gamma_m n_n + \\ \gamma_c \Gamma_{cd}(\delta_{dn} n_m + \delta_{dm} n_n) + (\gamma_a \gamma_a + 2 v n_m n_n + v^2) n_m n_n , \qquad (2.3)$$

respectively.

In the typical epitaxial systems, the matrix $\Gamma_{ab}$ does not depend on the film thickness and it is defined only by the crystallography factors. It can be changed, however, when temperature changes due to thermal expansions, and in some other processes. In our further analysis this misfit matrix is assumed fixed. Thus, in our further analysis the system is characterized by the thermodynamic parameters of the gaseous or liquid phase and of the parameters $\gamma_a$ and $v$.

We begin with the case in which the shear vector $\gamma_a$ vanishes. Then, the equations (2.2) and (2.3) can be simplified to read

$$u_{i,j} = \Gamma_{ab}\delta_{ai}\delta_{bj} + vn_j n_i$$
$$X_{mn} - x_{mn} = (\Gamma_{ab} + \Gamma_{ba})\delta_{am}\delta_{bn} + \Gamma_{ab}\Gamma_{ad}\delta_{mm}\delta_{dn} + (2v + v^2)n_m n_n, \quad (2.4)$$

respectively.

**The equations of equilibrium.** Let us consider an actual domain in the form of a straight cylinder with the base area $B_L$ and the fixed actual height $H$. The upper part $z$ of the cylinder is filled with the melt, whereas the remaining part $(H-z)$ is filled with the uniformly stressed solid. Let $\Sigma_l$ be the area of the cylinder base in the actual configuration, and $s_s$ be the area of the pre-image of the actual base in the reference configuration.

The following formulas are nothing more than the mass conservation equations:

$$\frac{M_l}{\Sigma_l \rho_l} + \frac{M_s}{S_s(1+v)m_s} = H, \quad M_s + M_l = M. \quad (2.5)$$

When minimizing the total energy (1.2), the constrains (2.5) should be taken into account. Introducing two indefinite Lagrange multipliers, we arrive at the following functional

$$\Psi_\Lambda = \Psi_\Lambda(M_s, M_l, v, \rho_l) =$$
$$M_s \psi_s(u_{i,j}) + M_l \psi_l(\rho_l) - \lambda_M (M_s + M_l) - \lambda_H \left( \frac{M_l}{\Sigma_l \rho_l} + \frac{M_s}{S_s m_s} \frac{1}{1+v} \right). \quad (2.6)$$

Straightforward calculation leads to the following formula of the first energy variation:

$$\delta\Psi_\Lambda = \left( \psi_s - \lambda_M - \lambda_H \frac{1}{S_s m_s} \frac{1}{1+v} \right)\delta M_s + \left( \psi_l(\rho_l) - \lambda_M - \lambda_H \frac{1}{\Sigma_l \rho_l} \right)\delta M_\rho +$$
$$M_l \left( \frac{\partial \psi_l(\rho_l)}{\partial \rho_l} + \lambda_H \frac{1}{\Sigma_l \rho_l^2} \right)\delta\rho_l + M_s \left( \frac{\partial \psi_s}{\partial u_{i,j}} n_i n_j + \lambda_H \frac{1}{S_s m_s} \frac{1}{(1+v)^2} \right)\delta v. \quad (2.7)$$

The relationship (2.7) leads to the following equilibrium equations:

$$\frac{\partial \Psi_\Lambda}{\partial M_s} = 0 \quad \rightarrow \quad \psi_s(u_{i,j}) - \lambda_M - \lambda_H \frac{1}{S_s m_s} \frac{1}{1+v} = 0,$$

$$\frac{\partial \Psi_\Lambda}{\partial M_l} = 0 \quad \rightarrow \quad \psi_l(\rho_l) - \lambda_M - \lambda_H \frac{1}{\Sigma_l \rho_l} = 0,$$

$$\frac{\partial \Psi_\Lambda}{\partial \rho_l} = 0 \quad \rightarrow \quad \frac{\partial \psi_l(\rho_l)}{\partial \rho_l} + \lambda_H \frac{1}{\Sigma_l \rho_l^2} = 0, \quad (2.8)$$

$$\frac{\partial \Psi_\Lambda}{\partial v} = 0 \quad \rightarrow \quad \frac{\partial \psi_s(u_{i,j})}{\partial u_{i,j}} n_i n_j + \lambda_H \frac{1}{S_s m_s (1+v)^2} = 0.$$

Eliminating the Lagrange multipliers $\lambda_M$ and $\lambda_H$, we arrive at the following two equilibrium equations:

$$S_s m_s (1+v)^2 \frac{\partial \psi_s}{\partial u_{i,j}} n_i n_j = \Sigma_l \rho_l^2 \frac{\partial \psi_l}{\partial \rho_l},$$

$$\psi_s - \lambda_H \frac{1}{S_s m_s} \frac{1}{1+v} = \psi_l - \lambda_H \frac{1}{\Sigma_l \rho_l}. \quad (2.9)$$

The first part of the equation (2.9) is mandatory for mechanical equilibrium between the crystal and the melt, whereas the second part is mandatory for equilibrium with respect to mass exchange between the phases.

**The stability conditions**. Let us turn now to the problem of equilibrium with respect to the Gibbs variations. The property of stability depends on the second variation of $\Psi_\Lambda$ in vicinity of equilibrium configuration. The required routine calculation leads us to the following formula:

$$\delta^2 \Psi_\Lambda = K_l (\delta \rho_l)^2 + K_s (\delta v)^2, \quad (2.10)$$

where the coefficients $K_l$ and $K_s$ are defined as

$$K_l = \frac{1}{2} M_l \left( \frac{\partial^2 \psi_l}{\partial \rho_l^2} + \frac{2}{\rho_l} \frac{\partial \psi_l}{\partial \rho_l} \right), \quad K_s = M_s \left( \frac{1}{2} \frac{\partial^2 \psi_s}{\partial u_{i,j} \partial u_{k,l}} n_i n_j n_k n_l + \frac{1}{1+v} \frac{\partial \psi_s}{\partial u_{i,j}} n_i n_j \right). \quad (2.11)$$

The bulk stability of any liquid phase implies the following universal thermodynamic inequality:

$$K_l \geq 0. \quad (2.12)$$

The bulk stability of any elastic phase – liquid or solid – demands the following thermodynamics inequality:

$$\frac{\partial^2 \psi_s}{\partial u_{i,j} \partial u_{k,l}} n_i e_j n_k e_l \geq 0, \tag{2.13}$$

for any two real vectors $n_i$ and $e_j$. That inequality should be satisfied in vicinity of each configuration.

In view of (2.13) the inequality

$$\frac{\partial^2 \psi_s}{\partial u_{i,j} \partial u_{k,l}} n_i n_j n_k n_l \geq 0 \tag{2.14}$$

is automatically implied by the general principles of thermodynamics for any anisotropic nonlinear elastic substance. The inequality (2.14) implies required inequality

$$K_s \geq 0 \tag{2.15}$$

only in vicinity of stress-free configuration. It still remains unknown whether or not the inequality (2.15) is mandatorily implied by thermodynamics in vicinity of any stable configuration of nonlinear elastic substance of any symmetry. However, all known experiments confirm the inequality (2.15) for any sufficiently small loadings.

The relationships (2.10)–(2.12), and (2.15) imply the stability of the heterogeneous system with respect to any Gibbs variations.

**The case of variable shift** $\gamma_a$. In this case the calculations become much more cumbersome but still manageable. The set of equations (2.8) should be amended with one more equilibrium condition:

$$M_s \frac{\partial \psi_s(u_{i,j})}{\partial u_{i,j}} \delta_{ai} n_j = 0. \tag{2.16}$$

Instead of (2.10) and (2.11), we arrive at the more general formula of the second energy variation also depending on $\delta \gamma_i$:

$$\delta^2 \Psi_\Lambda = C_l (\delta \rho_l)^2 + C_s (\delta v)^2 + C_{ik} \delta \gamma_i \delta \gamma_k, \tag{2.17}$$

where the coefficients $C_{ik}$ are given by the formula

$$C_{ik} = \frac{1}{2} M_s \frac{\partial^2 \psi_s(u_{i,j})}{\partial u_{i,j} \partial u_{k,l}} n_j n_l. \tag{2.18}$$

In view of the thermodynamic universal inequality (2.13) the last term $C_{ik} \delta \gamma_i \delta \gamma_k$ in formula (2.17) of the second variation $\delta^2 \Psi_\Lambda$ is positive. Thus, the statement regarding the stability of heterogeneous equilibrium configuration with respect to the Gibbs variations remain valid in this more general case.

## *The Universal Morphological Instability "Stressed Crystal – Melt"*

In order to understand the morphological instability "stressed crystal – melt" we should considerably widen the class of the Gibbs variations by including variations with non-uniform displacement gradients $\delta u_i$ and non-flat geometry of the phase interface. In this section, we dwell only on a brief qualitative discussion of this instability which has already become very popular in several branches of physics and engineering – a more detailed discussion can be found in [5–10].

**The bulk elastic energy balance**. Though the instabilities in question were established and will be discussed in the framework of continuum theory, it is easier to explain the idea rearrangement using the picture of a discrete lattice (array) of the atoms (elementary material particles). Below, we consider a two-dimensional model to simplify the picture. Figure 3 sketches a set of "atoms" having the shape of a rectangular plate with macroscopic sizes $L_x$ and $L_z$.

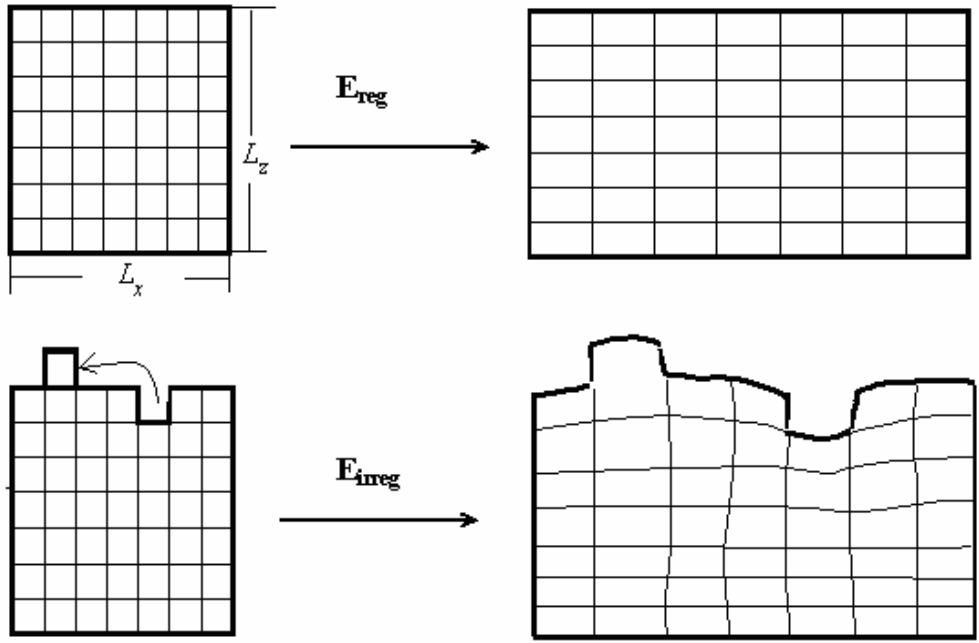

Figure 3. Surface corrugations of a free boundary diminishes an accumulated elastic energy of a nonhydrostatically stressed plate.

Assume now that the plate is subjected to prescribed uniform displacements at the vertical boundaries $L_z$ generating uniform horizontal stresses $P_{xx}$. At $P_{zz} = 0$, these displacements produce uniaxial compression with maximal shear stresses $\tau = 0.5 P_{xx}$ at the bisectional cross-sections. This deformation results in accumulating a certain amount of the elastic energy $E_{reg}$.

Now, imagine that migration of the "atoms" along the upper horizontal is allowed via the process of melting/crystallization; then, the phase interface no longer remains flat. In particular, we can consider the rearrangement of material particles of the crystal in which the total mass of the melt and crystal remain unchanged. For such variations the net change in the accumulated bulk energy is due to the change in the energy accumulated by the crystalline plate $E_{irreg}$. The calculations based on the exact nonlinear theory of anisotropic elasticity [5–7] imply the following key inequality:

$$E_{irreg} < E_{reg}, \tag{3.1}$$

regardless of the specific symmetry elastic modulae, etc. We will give in this paper several arguments of this unexpected statement which allows us to conclude that the flat interface of a solid is unstable with respect to mass rearrangement at any small stress.

It is a surprising fact because the irregularities of the boundary produce concentrators of elastic stresses and the local elastic energy density may tend to infinity in a small vicinity of the irregularities. (The magnitude of the stresses depends on the scale of "corrugation" – the shorter sizes of surface defects the higher the intensity of the local stresses). Thus, arbitrarily small shear stresses cause the instability of a flat smooth boundary with respect to the surface disturbances generated by the migration of "atoms" (in other words, with respect to rearrangement of the material particles). In the absence of surface tension, the flat boundaries permitting surface diffusion are unstable with respect to the disturbances of arbitrarily small tangential wavelengths. But, as it was also established in previous sections, the surface tension suppresses unstable growth of the irregularities with tangential wave lengths less than $\lambda_{crit}^{\sigma}$ given by the following formula:

$$\lambda_{crit}^{\sigma} = \frac{\pi\mu\sigma}{\tau^2}, \tag{3.2}$$

where $\sigma$ is the coefficient of surface tension, $\mu$ is the rigidity of solid (equal to shear elastic module in the case of isotropic elastic substance), and $\tau$ is the maximal shear stresses.

The combined action of stress and rearrangement destabilizes essentially internal interfaces separating two solids. The adjacent solid can suppress the stress driven rearrangement instability or ease it into making the phenomena even more dramatic – in any event, the set of possible scenarios is much richer in this case. The stability criteria now depend on the specific mechanisms of mass rearrangement and material properties of the adjacent solids. We demonstrate in this paper how these phenomena can be explored in different specific circumstances.

**An intuitive interpretation of the instability**. The stress driven rearrangement instability is a purely thermodynamic (energetic) phenomenon. Specific mechanisms of rearrangement of the particles — surface diffusion, vaporization-sublimation, melting-crystallization, migration of defects — play a secondary role in destabilization (defining the time-scale of the evolution but not the very fact of the occurrence of instability).

We explicitly consider two physical effects, namely, elasticity and surface energy. The stresses within the solid can be generated by an applied stress or be internal stresses, such as those associated with heteroepitaxy. For the sake of simplicity, we consider that the solid is only two-dimensional and assume that deposition takes place in the form of elementary square cells of material, as per Figure 4. We view each cell as a continuum so as not have to justify the application of elasticity and surface energy at an atomic scale. Uniaxial in-plane deformation changes the shape of the elementary cell. In particular, an elementary cell of the complete layers becomes a rectangle rather than a square. Thus, the material being deposited has a different lattice parameter than the substrate due to the presence of the uniaxial, lateral stress. When cell A attaches to the uniform ad-layer under it, its bottom stretches to match the lattice parameter of the strained ad-layer. Its top, on the other hand, remains at its initial unstrained width and the initially rectangular cell distorts into a trapezoidal shape. Consider now the possible locations for cell B to attach to the film in the vicinity of cell A. Particle B may attach itself to the ad-layer in, e.g., positions 1, 2, 3, or 4. Since surface energy favors as large a number of nearest neighbors as possible, sites 2 and 3 are preferable to 1 and 4 due to the proximity of cell A. This is why surface energy favors the growth of as smooth a surface as possible. If cell B attaches to site 1 or 4 it will take on the strained, trapezoidal shape of cell A. If, on the other hand, it attaches to site 2 or 3, the wall B shares with A becomes vertical and therefore both cells A and B become more strained than if cell B was at either site 1 or 4. Therefore, strain energy works against the surface smoothing tendencies of surface energy. Now, consider cell B becoming attached to site 5, on top of A. Since site 5 has the same number of nearest neighbors as site 1 and 4, the surface energy associated with B is the same for attachment to sites 1, 4 or 5. However, while the bottom of cell B would be stretched at sites a 1 or 4, its bottom is unstretched at site 5 because the top of cell A is unstretched. Therefore, consideration of strain energy favors site 5 over sites 1–4. Depending on the ratio of the surface energy to the strain energy, site 5 may (small ratio) or may not (large ratio) be favored over sites 2 or 3. If the strain and/or the elastic constant are large, the roughness of the surface will increase with continued film growth. This is the stress driven morphological instability.

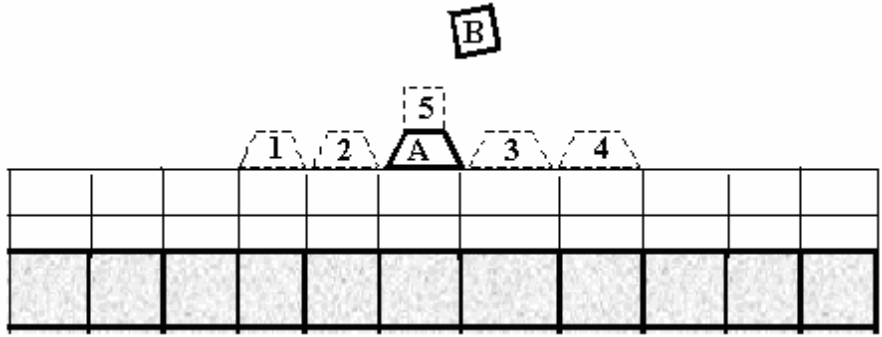

Figure 4. The mechanism of the stress driven rearrangement instability in epitaxial crystal growth.

This interpretation of the instability "stressed crystal – melt" was treated as the plausible mechanism of the Stranski–Krastanow or mixed pattern of epitaxial growth. This interpretation, however, still faces serious physical obstacles related to the nature of the van der Waals forces [13].

## *The Stress Driven Instabilities in Ceramic Armor*

The previously mentioned interpretations of the SDRI "stressed crystal – melt" clearly demonstrate the importance of the crystalline structure of the solid substance. However, there are some plausible manifestations of the SDRI in ceramics and amorphous glasses.

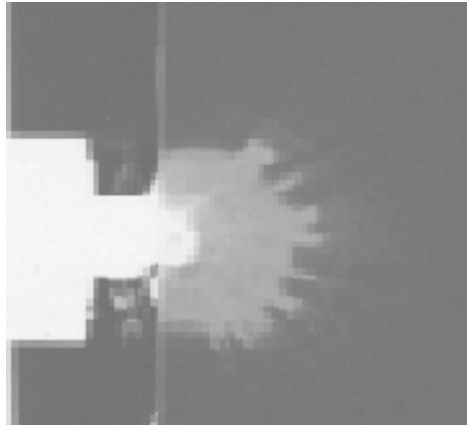

Figure 5. Morphology of failure fronts in the transparent aluminum oxinitride (courtesy of Dr. J. McCauley [15])

Various experiments on penetration of high-speed indenters into glasses and brittle ceramics brought experimenters to the conclusion of appearance of the so-called failure fronts. Failure fronts, so far detected, propagate with velocities which are smaller than the velocities of bulk acoustic waves. Particularly convincing evidence of existence of failure fronts in glasses and in transparent ceramics is delivered by ultra-fast video cameras making $10^6$–$10^7$ shots per second. These experiments have demonstrated that the failure fronts are quite rough [15,16]. Especially impressive photographs have been obtained in the experiments on aluminum oxinitride shown in Figure 5. One can clearly see in this figure the appearance of spikes of damaged substance with the tendency of their further growth rather than disappearance. This fact prompts one to think that the originally smooth failure fronts are morphologically unstable. This idea was converted into a quantitative theory in one paper [17], in which the intact and damage substances are treated as two different phases of the same material.

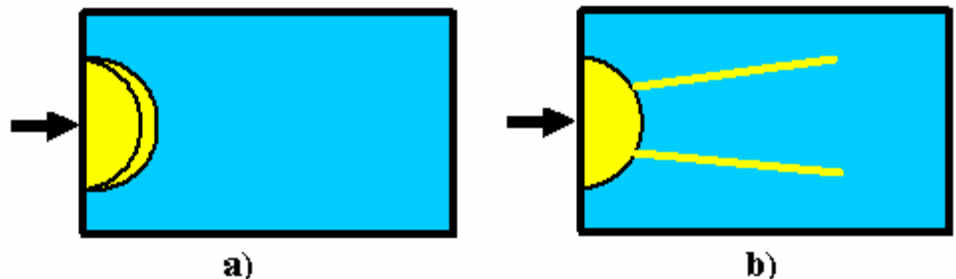

Figure 6. Towards intuitive interpretation of the SDRI of failure fronts.

Qualitatively, the appearance of the spikes can be explained as follows. The two sketches in Figure 6 show two different morphologies of the advancing flat. The front is originally of a smooth and circular shape. Figure 6a shows an advancing failure front that does not change its morphology and remains smooth and quasi-circular, and Figure 6b shows an advancing front in the form of two fingers. It is assumed that the two newly created comminuted areas on 6a and 6b located to the right from the original front are equal. It should be intuitively clear that the finger-like morphology of the failure front allows the release of much more elastic energy than the morphologically stable front shown in 6a. Indeed, the propagating front in 6a releases elastic energy only within the small domain between the original and final front positions, whereas the finger-like front allows the release of the accumulated elastic energy from everywhere to the right of the original front, including the material that basically remains in the intact state. So, the "fingering" of the front allows the release of the accumulated energy in a more efficient way than the morphologically stable propagation of the failure front.

## *Conclusion*

The classical Gibbs analysis [1] of equilibrium configurations of the heterogeneous systems "stressed crystal – melt" is extended further to include stability issues. As it was demonstrated two decades ago [5-7], the natural generalization of the original Gibbs methodology leads to the conclusion that phase interfaces in heterogeneous systems with stressed phase are morphologically unstable. In particular, it was demonstrated that the phase interfaces separating any nonhydrostatically stressed crystal and its melt is always unstable. We discussed several plausible manifestations of this universal instability related to epitaxial growth and fracture.

At the same time, in the vast majority of the cases neither the precise laboratory experiments nor the widespread engineering systems show the tendency of this sort of morphological destabilization. This absence of the stress driven destabilization requires a clear scientific interpretation which seems crucial for further developments in the Gibbs thermodynamics and in several physical and engineering disciplines. To that end we investigated the stability of the systems "stressed crystal – melt" with respect to the special class of the piece-wise uniform variations introduced by Gibbs. It is demonstrated that the equilibrium configurations "stressed crystal – melt" are stable with respect to the Gibbs variations. The established fact sheds some light on the (in)stability in the various systems "stressed crystal – melt" and can be used as the first step in developing the adequate theory of equilibrium and stability of those widespread systems.